%% file: preprint3.tex
\documentstyle[12pt,aaspp4]{article}

\newcommand{\etal}{~et~al.}

\newcommand{\suph}{\wisk{^{\rm h}}}
\newcommand{\supm}{\wisk{^{\rm m}}}
\newcommand{\sups}{\wisk{^{\rm s}}}
\newcommand{\wisk}[1]{\ifmmode{#1}\else{$#1$}\fi}

\newcommand{\dmr}{{\sl COBE}/DMR}
\newcommand{\suzie}{SuZIE}

\newcommand{\bfC}{\wisk{\mbox{\boldmath $C$}}}

\newcommand{\bfDelta}{\wisk{\mbox{\boldmath $\Delta$}}}
\newcommand{\bff}{\wisk{\mbox{\boldmath $f$}}}

\newcommand{\bfM}{\wisk{\mbox{\boldmath $M$}}}
\newcommand{\bfS}{\wisk{\mbox{\boldmath $S$}}}

\newcommand{\decoff}{\wisk{2^\prime}}
\newcommand{\fwhm}{\wisk{1.7^\prime}}
\newcommand{\throw}{\wisk{2.3^\prime}}

\begin{document}

\begin{flushright}
MIT-CTP-2567, KUNS-1412
\end{flushright}

\bigskip

\title{Using SuZIE arcminute-scale CMB anisotropy data to probe open
  and flat-$\Lambda$ CDM cosmogonies}

\author{
  K.~Ganga\altaffilmark{1,2},
  Bharat~Ratra\altaffilmark{3}, 
  S.E.~Church\altaffilmark{1}, 
  Naoshi~Sugiyama\altaffilmark{4},
  P.A.R.~Ade\altaffilmark{5},
  W.L.~Holzapfel\altaffilmark{1,6},
  A.E.~Lange\altaffilmark{1},
  and 
  P.D.~Mauskopf\altaffilmark{1,7}
  }

\altaffiltext{1}{Division of Physics, Mathematics and Astronomy,
                 Mail Stop: 59--33,
                 California Institute of Technology,
                 Pasadena, CA \ 91125.}
\altaffiltext{2}{Current address: IPAC, 
                                  California Institute of Technology, 
                                  MS 100--22,
                                  Pasadena, CA \ 91125}    
\altaffiltext{3}{Center for Theoretical Physics,
                 Massachusetts Institute of Technology,
                 Cambridge, MA \ 02139.
                 Current address:
                 Department of Physics,
                 Kansas State University,
                 Manhattan, KS \ 66506.}
\altaffiltext{4}{Department of Physics,
                 Kyoto University, 
                 Kitashirakawa-Oiwakecho,
                 Sakyo-ku, Kyoto 606,
                 Japan.}
\altaffiltext{5}{Department of Physics,
                 Queen Mary and Westfield College,
                 Mile End Road,
                 London, E1~4NS, UK.}
\altaffiltext{6}{Current address:
                 Laboratory for Astrophysics and Space Research,
                 Enrico Fermi Institute,
                 University of Chicago,
                 5641 S. Ingleside Ave.,
                 Chicago, IL \ 60637.}
\altaffiltext{7}{Current address: 
                 Department of Physics and Astronomy,
                 University of Massachusetts at Amherst,
                 Amherst, MA \ 01003--0120.}

\slugcomment{Submitted for publication in {\it The Astrophysical Journal}}

\begin{abstract}
  We use arcminute-scale data from the Sunyaev-Zel'dovich Infrared
  Experiment to set limits on anisotropies in the cosmic microwave
  background radiation in open and spatially-flat-$\Lambda$ cold dark
  matter cosmogonies. There are no 2-$\sigma$ detections for the
  models tested. The upper limits obtained are consistent with the
  amplitude of anisotropy detected by the {\sl COBE}/DMR experiment.
\end{abstract}


\section{Introduction}

With the detection of anisotropy in the cosmic microwave background
(CMB) by the {\sl COBE}/DMR experiment
(Smoot\etal~1992\markcite{Smoot}; Wright\etal~1992\markcite{Wright};
Bennett\etal~1996\markcite{Bennett}) and a host of other detections on
angular scales greater than $\sim 10'$
(Ganga\etal~1994\markcite{Ganga}; Hancock\etal~1996\markcite{Hancock};
Piccirillo\etal~1997\markcite{Piccirillo};
Netterfield\etal~1997\markcite{Netterfield};
Gundersen\etal~1995\markcite{Gundersen};
Tucker\etal~1997\markcite{Tucker}; Platt\etal~1996\markcite{Platt};
Masi\etal~1996\markcite{Masi}; Tanaka\etal~1996\markcite{Tanaka};
Inman\etal~1996\markcite{Inman}; Griffin\etal~1997\markcite{Griffin};
Scott\etal~1996\markcite{Scott}), CMB anisotropy measurements have
become a powerful tool for testing models of cosmic structure
formation.  At smaller angular scales, however, the detection of
anisotropies has proven more difficult
(Griffin\etal~1997\markcite{Griffin}; Myers\etal~1993\markcite{Myers};
Subrahmanyan\etal~1993\markcite{Subrahmanyan}). Moreover, calculation
of theoretical CMB anisotropies at these angular scales is difficult
and has only recently been done accurately. Thus, the detection of
arcminute scale anisotropies and their classification is one of the
significant hurdles to be overcome in the characterization of the CMB.

In this paper we use arcminute-scale CMB anisotropy data from the
Sunyaev-Zel'dovich Infrared Experiment~(\suzie), an experiment
designed to measure the Sunyaev-Zel'dovich (SZ) effect in distant
clusters of galaxies. These data were taken as part of a program to
characterize the instrument for SZ observations, but can be used to
set interesting limits on the power spectrum of CMB anisotropies.
Theoretical predictions for primary CMB anisotropy spectra are used in
conjunction with these data to set limits on the normalization of
cosmic structure formation models in low-density open and
spatially-flat (with a cosmological constant $\Lambda$) cold dark
matter (CDM) universes.

\section{Data}

\suzie\ was originally conceived and constructed to make
millimeter-wave observations of both the thermal and kinetic SZ
effects in clusters of galaxies from the Caltech Submillimeter
Observatory on Mauna Kea (Holzapfel\etal~1997a\markcite{Holzapfel97a};
Wilbanks\etal~1994\markcite{Wilbanks2}). It employs a 2$\times$3 array
of bolometers cooled to 300~mK, with beam widths of
approximately~\fwhm\ (FWHM) and beam throws of~$\pm\throw$ in right
ascension (see insets of Figure~\ref{fig:dataplot}; also Figure~2 of
Church\etal~1997\markcite{Church97}). The instrument is described
fully by Holzapfel\etal~(1997b)\markcite{Holzapfel97b}.

Each pair of detectors in a single row is electronically differenced
in an AC bridge, with the output of each bridge phase-synchronously
demodulated to produce a DC signal proportional to the difference in
power absorbed by the two detectors~(Rieke\etal~1989\markcite{Rieke};
Wilbanks\etal~1990\markcite{Wilbanks3}; Glezer, Lange, \&
Wilbanks~1992\markcite{Glezer92}). In terms of atmospheric rejection,
this is equivalent to a square wave chop between the two detectors at
infinite frequency.  Observations were made at 140~GHz ($\sim$~11\%
bandpass) and were performed by fixing the telescope and using the
rotation of the Earth to make the sky `drift' (in right ascension)
through the array's view. This eliminates spurious signals arising
from mechanical chopping or motion of the telescope.

The output from the AC bridge, after calibration, gives the difference
in temperature between two spots on the sky,
\begin{equation}
  \Delta_{ij} = T_{i} - T_{j},
\end{equation}
where $T_i$ and $T_j$ are the CMB temperature at the two spots on the
sky.  The focal plane has two rows of three pixels. The rows are
separated by $2.0'$ in declination, with the three pixels in each row
separated by $2.3'$ in right ascension. The data used here are the
differences between the two outermost pixels in each row, detectors 3
and 1 in row 1 and detectors 6 and 4 in row 2, which are separated by
$4.6'$ (Figure~1; also see Figure~2 of
Church\etal~1997\markcite{Church97}).

Two $36'$ regions of the sky were observed, one centered at R.A.
10\suph~21\supm~49.00\sups\ and Dec.
4$^\circ$~4$^\prime$~23.00$^{\prime\prime}$ (Region 1) and the other
at R.A. 16\suph~30\supm~17.00\sups\ and Dec.
5$^\circ$~56$^\prime$~0.00$^{\prime\prime}$ (Region 2; epoch 1950).
The instrument is Nyquist sampled at 5~Hz.  The data have an offset
and linear drift removed, and are coadded and binned into
45$^{\prime\prime}$ bins. The data are shown in
Figure~\ref{fig:dataplot} and the zero-lag window function associated
with one row of pixels, $W_l$, is shown in Figure~\ref{fig:win_mod}.
The uncertainty in the gaussian beam widths, estimated at 5\%,
translate to errors in limits on models primarily through the
uncertainty introduced into the calibration. The beam width
uncertainties are ignored otherwise (Ganga et al. 1997). The
calibration, obtained from scans of Uranus, is known to 8\%. The data
reduction and calibration are described more fully
in~Church\etal~(1997)\markcite{Church97}.

\placefigure{fig:dataplot}

\placefigure{fig:win_mod}

\section{Models}

The CMB fractional temperature perturbation, $\delta T/T$, can be
decomposed into a sum over spherical harmonics according to
\begin{equation}
  {\delta T \over T}(\theta , \phi) = \sum_{l=2}^\infty
  \sum_{m=-l}^l a_{lm} Y_{lm}(\theta, \phi).
\end{equation}
In this scheme, a convenient characterization of the CMB anisotropy is
its angular power spectrum, $C_l$, defined in terms of the ensemble
average
\begin{equation}
  \langle a_{lm} a_{l^\prime m^\prime}^* \rangle = C_l 
  \delta_{ll^\prime} \delta_{mm^\prime}.
\end{equation}

Prior to the construction of semi-realistic cosmogonies, it was
conventional to use a gaussian parameterization for the CMB anisotropy
spectrum.  The fiducial CDM (fCDM) model, an Einstein-de Sitter
cosmological model with gaussian, adiabatic, scale-invariant
energy-density perturbations (Harrison 1970\markcite{Harrison};
Peebles \&\ Yu 1970\markcite{Peebles70}; Zel'dovich 1972\markcite{Zeldovich}),
focussed attention on what has now come to be known as the flat angular 
spectrum (Peebles 1982\markcite{Peebles82}),
\begin{equation}
  C_l = {6C_2 \over l(l+1)} = {24 \pi \over 5} {(Q/T_0)^2 \over l(l+1)}.
\end{equation}
In this context, $Q$, sometimes referred to as $Q_{\rm rms-PS}$, is
the {\em implied} quadrupole amplitude of the CMB anisotropy spectrum
and $T_0$ is the present CMB temperature. $Q$ should not be confused
with the actual amplitude of the quadrupole moment of the CMB anisotropy, 
to which SuZIE is not sensitive. This spectrum is shown as the straight, 
solid line in Figure~\ref{fig:win_mod}.

The flat spectrum is simply a large-scale approximation of the fCDM
spectrum. On smaller scales, the pressure of the photon-baryon fluid
causes the spectrum to deviate from this approximation. The fCDM
spectrum, with standard recombination, $h = 0.5$, and baryonic-mass
density parameter $\Omega_B = 0.0125h^{-2}$, is shown as the dotted
line in Figure~\ref{fig:win_mod}. Here, the Hubble parameter $H_0 =
100h$ km s$^{-1}$ Mpc$^{-1}$. Although still of some historical
interest, the fCDM model, which has a clustered-mass density parameter
$\Omega_0 = 1$, is now known to be an inadequate representation of the
observed universe.

Low-density CDM cosmogonies, with flat spatial sections and a
cosmological constant or with open spatial sections and no $\Lambda$,
are consistent with a large fraction of present observations.  For
flat-$\Lambda$ models, see Stompor, G\'orski, \& Banday
(1995)\markcite{Stompor}, Ostriker \& Steinhardt
(1995)\markcite{Ostriker}, Ratra et al. (1997)\markcite{Ratra97},
Liddle et al. (1996)\markcite{Liddle}, and Ganga, Ratra, \& Sugiyama
(1996)\markcite{GRS}, and for open models see Kamionkowski et al.
(1994)\markcite{Kamionkowski}, Ratra et al.
(1997)\markcite{Ratra97}, Ganga et al.~(1996)\markcite{GRS}, and
G\'orski et al. (1996)\markcite{Gorski96}.  Since the low-density
$\Lambda$-CDM models are assumed to have flat spatial sections, the
simplest power spectrum for gaussian, adiabatic, energy-density
perturbations is the scale-invariant one.  The simplest spectrum for
these perturbations consistent with open spatial sections is that
which is generated by quantum-mechanical fluctuations during an early
epoch of inflation in an open model (Ratra \& Peebles
1994\markcite{Ratra94}, 1995\markcite{Ratra95}; Bucher, Goldhaber, \&
Turok 1995\markcite{Bucher95}; Yamamoto, Sasaki, \& Tanaka
1995\markcite{Yamamoto}).  The values of the parameters $\Omega_0$,
$h$ and $\Omega_B$ which characterize the spectra in the low-density
CDM models used here are chosen to be roughly consistent with present
observational estimates of $\Omega_0$, $h$, the age of the universe,
and the constraints on $\Omega_B$ that follow from the observed light
element abundances in the standard nucleosynthesis model (Ratra et
al.~1997\markcite{Ratra97}; see Table~1 for parameter values). A
standard recombination open model spectrum and a standard
recombination flat-$\Lambda$ model spectrum are shown in
Figure~\ref{fig:win_mod}.

The general procedure for the computation of CMB anisotropy spectra
is discussed by Sugiyama (1995)\markcite{Sugiyama}.  Since
SuZIE is sensitive to sub-arcminute-scale anisotropies, we have
computed the open model spectra to $l = 7500$, where photon diffusion
strongly suppresses the power. Photon diffusion cuts off power on
larger scales in the fiducial and flat-$\Lambda$ CDM models, so we
have only computed the spectra in these models up to $l = 3500$.

\section{Analysis}

As a starting point for our analysis, we note that Church et al.
(1997) have shown that the reduced $\chi^2$'s for each of the data
sets used here are less than one. This gives us a good indication that
the data does not show a detection; i.e., it is consistent with
experimental noise.  As shown below, this is indeed the case. To set
numerical limits on various cosmological models, however, and in order
to compare the results obtained for different models, we use a more
refined technique.

A primary goal of CMB anisotropy observations is to measure
the angular spectrum of the fluctuations. Given the width of the SuZIE
window function, it is not possible to extract a value for each
multipole moment to which the experiment is sensitive.  Rather, one
can hope, at best, to determine a few observational numbers with
reasonable accuracy.  It is therefore necessary to simplify the
problem by assuming a model for the spectrum over the range of $l$ to
which the experiment is sensitive.  Typically, one assumes a
functional form for the shape of the spectrum as a function of $l$,
with the overall normalization allowed to be a free parameter. The
shape of the spectrum can also depend on other free parameters. One
then compares this family of spectra to the data and determines the
value of the normalization and other parameters which best reproduce
the data.

For the purpose of the following discussion we assume that the SuZIE
sky signal is purely CMB without any offset or gradient, which would
have been removed during data reduction.  Following Bond et
al.~(1991)\markcite{Bond91} to account for the baseline removal, the
likelihood, at the nominal calibration, corresponding to a given
spectrum is
\begin{equation}
  L \propto 
  {1 \over \sqrt{\det(\bff^T \bfM^{-1} \bff) \det(\bfM)}}e^{-\chi^2/2},
\label{eq:likeqn}\end{equation}
where,
\begin{equation}
  \chi^2 = \bfDelta^T \Bigl(\bfM^{-1} -
  \bfM^{-1}\bff ( \bff^T\bfM^{-1}\bff)^{-1}\bff^T{\bfM^{-1}}^T \Bigr)
  \bfDelta .
\label{eq:chieqn}\end{equation}
Here, $\bfDelta$ is the vector of temperature differences and $\bfM$
is the correlation matrix of the data in the context of the model
considered.  The correlation matrix, \bfM, is the sum of correlations
arising from experimental noise, \bfS, and those from correlations
intrinsic to the CMB, \bfC. That is, $\bfM = \bfC + \bfS$. In the case
of \suzie, the noise correlation matrix is approximately diagonal,
with the diagonal elements simply equal to the variance of the
corresponding measurement. This is true so long as atmospheric
correlations and other correlations induced in the process of reducing
the data can be ignored (Church et al. 1997). \bfC\ is calculated
according to
\begin{equation}
  C_{ij} = {1\over 4\pi} \sum_l (2l+1)C_l W_{ijl}, 
\end{equation}
where 
\begin{equation}
  W_{ijl} = {16\pi \over 2l+1} B_{il} B_{jl} \sum_{m=1}^l
        \left[{\sin \left(m\Phi_b/2\right) \over m\Phi_b/2}\right]^2
        \left[1-\cos\left(m\Phi_0\right)\right]
        \hat{P}_{lm}\left(\theta_i\right)\hat{P}_{lm}\left(\theta_j\right)
        \cos\left[m\left(\phi_i-\phi_j\right)\right].
\label{eq:fullwin}\end{equation}
Here $B_{il} = \exp[-l(l+1) \sigma_i^2/2]$ is the Legendre transform
of the $i^{\rm th}$ beam, assumed to be gaussian, $\Phi_b$ is the
extent of a single data bin in right ascension,
$0.75'\times\cos\delta$ in this case, and $\Phi_0$ is the azimuthal
separation of the two points differenced in a single sample, $4.6'$.
$\theta$ is the declination of the pixel above or below the {\em
  center} of the array ($\pm 1'$), and $\hat{P}_{lm}$ is defined by
\begin{equation}
  Y_{lm}(\theta, \phi) = \hat{P}_{lm}(\theta)\cdot e^{im\phi}. 
\end{equation}
See Bond (1996) and White \& Srednicki (1995) for general discussions
of CMB window functions. In eqs.~(\ref{eq:likeqn})
and~(\ref{eq:chieqn}), $\bff$ is the array of functions that have been
fit out of the data (in this case an offset and gradient from each
channel; it is thus an $96 \times 4$ matrix for each region).

With this prescription, all models yield the same likelihood at $Q =
0~\mu$K (i.e., nothing but noise), providing a convenient way to
normalize across models.

Following Ganga et al.~(1997), we account for calibration uncertainty by
taking 
\begin{equation}
  {\cal L} = {1 \over \sqrt{2\pi} \, \sigma_C \, Q}
  \times \int_0^\infty dQ' \, e^{-(Q' - Q)^2
    /[2(\sigma_C Q)^2]} L(Q'),
\end{equation}
where $\sigma_C$ is the fractional uncertainty in the calibration,
0.08.  This is conservative since it assumes the calibration
uncertainty is completely correlated in all detectors. This is the
case if there is an error in the model of the calibration source, but
does not apply if the uncertainties are statistical, due to, for
example, uncertainty in the beam measurements.

This calibration-uncertainty-corrected likelihood is computed for all
the models considered. Assuming a uniform prior in $Q$ gives a
posterior probability density distribution equal to the likelihood
function.

For the models considered here there are no 2-$\sigma$ anisotropy
detections, defined according to the prescription discussed in Ganga
et al.~(1997; also see Berger~1985). To quantify the constraints, we
quote two different 2-$\sigma$ upper limits for each model and data
set considered. The highest posterior density (HPD) prescription
provides the smallest possible 2-$\sigma$ upper limit. In this scheme,
the upper and lower 2-$\sigma$ limits on $Q$, $Q_u$ and $Q_l$, are
defined such that
\begin{equation}
  \int_{Q_l}^{Q_u} {\cal L} \cdot {\rm d}Q
  = 0.9545\times\int_{0}^{\infty}{\cal L} \cdot {\rm d}Q,
\label{eq:hpd}\end{equation} 
and $Q_u - Q_l$ is {\em minimized}. Here, $Q_l$ is zero for all
models, indicating that there are no detections. The equal tail (ET)
prescription gives a larger 2-$\sigma$ upper limit, calculated using
\begin{equation}
  \int_0^{Q_u} {\cal L} \cdot {\rm d}Q
  = 0.9772\times\int_{0}^{\infty}{\cal L} \cdot {\rm d}Q.
\label{eq:eqt}\end{equation}

Finally, we note that expected sky rms can be written as a sum of
contributions from different multipoles, $\delta T_{\rm rms} {}^2 =
\sum_l (\delta T_{\rm rms} {}^2)_l$, where
\begin{equation}
  (\delta T_{\rm rms} {}^2)_l = T_0^2 {(2l + 1) \over 4 \pi} C_l W_l.
\label{eq:dtrms}\end{equation}
In Figure~\ref{fig:trms} we plot this quantity for the four
representative spectra shown in Figure~\ref{fig:win_mod}. This plot
indicates that though the SuZIE window function peaks at $l \sim
2400$, with $l_{\rm eff} \sim 2340$ and $l_{e^{-0.5}} \sim 1330$ and
3670 (as defined by Bond 1996), the SuZIE sensitivity is model
dependent. In these models SuZIE is most sensitive to multipoles $l
\lesssim 1000$, and is still sensitive down to $l \sim 200$.

\placefigure{fig:trms}

\section{Results and Discussion}

Table~1 lists the 2-$\sigma$ upper limits on $Q$. Also listed are the
limits on the bandtemperature, $\delta T_l$, obtained using the upper
limits on $Q$, equation~(\ref{eq:dtrms}), and
\begin{equation}
  \delta T_l = 
  \sqrt{
    \delta T_{\rm rms} {}^2  \over
    \sum_{l=2}^\infty
    \left[
      { \left(l+1/2\right)W_l \over
        l\left(l+1\right) }
    \right]
  }.
\label{eq:tleqn}\end{equation}
Note that $\delta T_l$ is {\em not} a function of $l$ (Bond 1996, also
see Ganga et al.~1997, eq. [7]). Like $\delta T_{\rm rms}$, $\delta
T_l$ should be fairly independent of the model, but unlike $\delta
T_{\rm rms}$, it can also be compared to results from experiments with
different window functions.  The degree to which $\delta T_l$ varies
from model to model gives an indication of how good the flat bandpower
approximation is.

Figure~\ref{fig:likeplot} shows representative likelihood functions
for the combined SuZIE data and some of the models considered.


\placefigure{fig:likeplot}
 
In all cases, the SuZIE 2-$\sigma$ upper limits on $Q$ are consistent
with, and less restrictive than, the detections obtained by the \dmr\ 
experiment. In general, these limits are lower for the open models
than for the flat-$\Lambda$ models.  The 2-$\sigma$ upper limits on
$\delta T_l$ are much less model dependent than the $Q$ limits. There
remains, however, a $\sim 25\%$ difference in $\delta T_l$ between the
two extreme cases, indicating that using the flat bandpower
approximation to calculate the expected signal that should be seen by
an experiment is not accurate at these angular scales.

Like open CDM cosmogonies, anti-tilted flat-$\Lambda$ models ($n > 1$)
predict more power on arcminute angular scales than do flat-$\Lambda$
models without tilt ($n = 1$). The SuZIE 2-$\sigma$ upper limits on
$Q$ for models with $\Omega_0 = 0.3$, $h = 0.6$, $\Omega_Bh^2 =
0.0175$ and $n = 1.2$ and 1.4 are $18~\mu K$ and $11~\mu K$ (HPD),
respectively.  These can be compared to the limits for the
twenty-first model of Table~1, which has $n=1$.  When these results
are combined with the COBE/{\sl DMR} result, it is likely that
spectral indices of order 1.5 or larger can be ruled out for this
particular cosmogony.

Atmospheric noise is a concern for all ground based CMB anisotropy
experiments. Because we have assumed that the experimental noise is
gaussian, if it were to be significantly non-gaussian it could change
the limits set here. In order to check this, we have examined
histograms of the data in each bin and found them sufficiently
gaussian. We have also repeated the above analysis including estimates
of the correlations induced by the atmosphere (Church et al. 1997). In
all cases, these limits are comparable to or less than those quoted
above without the atmospheric corrections. Contributions from dust at
these angular scales are not expected to be significant (Tegmark \&
Efstathiou 1996).

Since the telescope is generally stationary while data is taken, SuZIE
is mostly immune to problems caused by motion of the instrument.
Between scans, however, the telescope slews, raising the possibility
of temperature drifts in the instrument at the beginning of a new
scan. To check that this is not an issue, we repeated the above
analysis excluding the first seven bins in each scan. The results were
qualitatively the same as those from the entire data set, with limits
a bit less strict, numerically consistent with the loss of data.
Church et al. (1997)\markcite{Church97} have done a more sophisticated
analysis by correlating the data with the temperature of the \suzie\ 
300~mK stage and come to the same conclusion.

We have checked that the truncated spectra used here does not affect
the results by ``filling in'' the $C_l$ spectra to $l = 10000$ using
the $C_l$ value at the largest computed $l$ value. As one might expect
from an inspection of Figure~\ref{fig:trms}, the numbers do not change
to the accuracy quoted here.

\section{Conclusion}

The SuZIE has been used to set limits on primary anisotropies in the
CMB at arcminute angular scales. Though there are no detections in
this data set, the likelihood functions shown in
Figure~\ref{fig:likeplot} peak tantalizingly close to the maximum
likelihood $Q$ values obtained from the \dmr\ data, lending hope that
CMB fluctuations will soon be detected on arcminute scales. Even if
they are not, however, increased sensitivity may allow for meaningful
limits to be set on cosmological parameters such as $\Omega_0$ or the
spectral index of primordial fluctuations.

The next generation version of the SuZIE instrument, SuZIE 1.5, is now
operational and has taken data simultaneously in three wavelength
bands: 1.1, 1.4, and 2.1~mm (Mauskopf et al.~1997\markcite{Mauskopf}).
These three bands correspond to three atmospheric windows near the
peak brightness of the CMB and close to the minimum in the confusion
expected from astrophysical sources on these angular scales (Fischer
\& Lange~1993\markcite{Fischer93}). The new receiver's ability to
observe simultaneously in these three bands will allow residual
atmospheric noise, which limited the sensitivity of the previous
instrument, to be identified by its frequency correlation between
channels and then removed.

\bigskip

We thank Thor Wilbanks for contributions to the SuZIE instrument and
for help with the observations. This work was supported in part by
NASA grant NAGW-4623 and by National Science Foundation grant
AST--95--03226. BR was supported in part by National Science
Foundation grant EPS-9550487 and matching support from the state of
Kansas. NS was supported by Grant-in-aid for Scientific Research No.
8740193 from the Ministry of Education of Japan. The CSO is operated
by the California Institute of Technology under funding from the
National Science Foundation, contract AST--93--13929.

\clearpage

\begin{table}[ht]
  \begin{center}
    \label{tab:limstab} 
    \caption{SuZIE Region 1, 2, and Combined 2-$\sigma$ Upper Limits on 
      $Q$ and $\delta T_l$ (in $\mu$K) for Various
      Models\tablenotemark{a}} \tablenotetext{a}{The first 13 models
      are open CDM, the 14$^{\rm th}$ is fCDM, the 11 following fCDM
      are flat-$\Lambda$ CDM models and, finally, the flat spectrum
      model is at the bottom. The quoted values of $\delta T_l$ are
      determined using the data from both Regions 1 \& 2.}
    \tablenotetext{b}{The HPD prescription is described in
      eq.~[\ref{eq:hpd}].} \tablenotetext{c}{The EQ prescription is
      described in eq.~[\ref{eq:eqt}].}
    \begin{tabular}{ccc|ccc|ccc|cc}
      \null \\
      \tableline\tableline
      \multicolumn{3}{c|}{ } &
      \multicolumn{3}{c|}{$Q$ (HPD)\tablenotemark{b}} &
      \multicolumn{3}{c|}{$Q$ (ET)\tablenotemark{c}} &
      \multicolumn{2}{c}{$\delta T_l$} \\
      \tableline
      $\Omega_0$ & $h$ & $\Omega_B h^2$ &
      R. 1 & R. 2 & R. 1 \& 2 & 
      R. 1 & R. 2 & R. 1 \& 2 & HPD\tablenotemark{b} & ET\tablenotemark{c} \\
      \tableline
      \input{table1.tex}
      \tableline
    \end{tabular}
  \end{center}
\end{table}


\clearpage
\centerline{\bf Figure Captions}

\begin{figure}[ht]
  \caption{Measured temperature differences on the sky as a function of
    scan position. The upper two panels show data from the two
    detector rows for Region~1, the lower two panels show data from
    the two detector rows for Region~2. As stated in the text, each of
    the four data sets correspond to the differences between two of
    the six pixels, as represented schematically by the six circles in
    each plot. Each pixel is separated from it's neighbors by~\throw\ 
    in right ascension and~\decoff\ in declination. The filled circles
    indicate the pixels being differenced for each data plot. See
    Church et~al.~(1997) for a full description of the data
    reduction.}
  \label{fig:dataplot}
\end{figure}
\begin{figure}[ht]
  \caption{SuZIE zero-lag window function ($W_l \equiv W_{iil}$ 
    in eq.~[\ref{eq:fullwin}]; scale on right axis) and four primary
    CMB anisotropy spectra normalized to the DMR two-year data (scale
    on left axis). The open, flat-$\Lambda$ and fCDM spectra are
    computed for $\Omega_B h^2 = 0.0125$. The open CDM, fCDM and flat
    spectra are normalized to the DMR galactic-frame maps (G\'orski et
    al. 1994, 1995; also see Ratra et al. 1997), the flat-$\Lambda$
    spectrum is normalized to the DMR ecliptic-frame maps (Bunn \&
    Sugiyama 1995; also see Ratra et al. 1997), and in all cases
    the quadrupole moment is excluded from the analysis of the DMR
    data. Since the DMR four-year data have been completely analyzed
    for only some of the models considered here, we use the DMR
    two-year data normalizations, bearing in mind that the four-year
    data indicates values for $Q$ which are $\sim 10\%$ smaller than
    those obtained with the two-year data (e.g., G\'orski et al.~1995,
    1996).  The flat bandpower 2-$\sigma$ (HPD) upper limit obtained
    here is placed at the location of the window function peak. See,
    however, Figure~\ref{fig:trms}.}
  \label{fig:win_mod}
\end{figure}
\begin{figure}[ht]
  \caption{$(\delta T_{\rm rms}{}^2)_l$ (eq.~[\ref{eq:tleqn}]),
    as a function of $l$ for the CMB anisotropy spectra shown in
    Figure~\ref{fig:win_mod}. These curves should be compared to the
    SuZIE window function shown in Figure~\ref{fig:win_mod}. Note the
    multiple ``sensitivity" peaks for each spectrum, and that the
    major contributions to the signals arise from multipole values
    lower than those indicated by considering the window
    function~(Figure 2) alone.}
  \label{fig:trms}
\end{figure}
\begin{figure}[hb]
  \caption{Likelihood functions given the combined Regions 1 and 2 SuZIE 
    data, for the four
    CMB anisotropy angular spectra of Figure~\ref{fig:win_mod}.}
  \label{fig:likeplot}
\end{figure}

\clearpage
\pagestyle{empty}

\begin{center}
  \leavevmode
  \epsfxsize=5.5truein
  \epsfbox{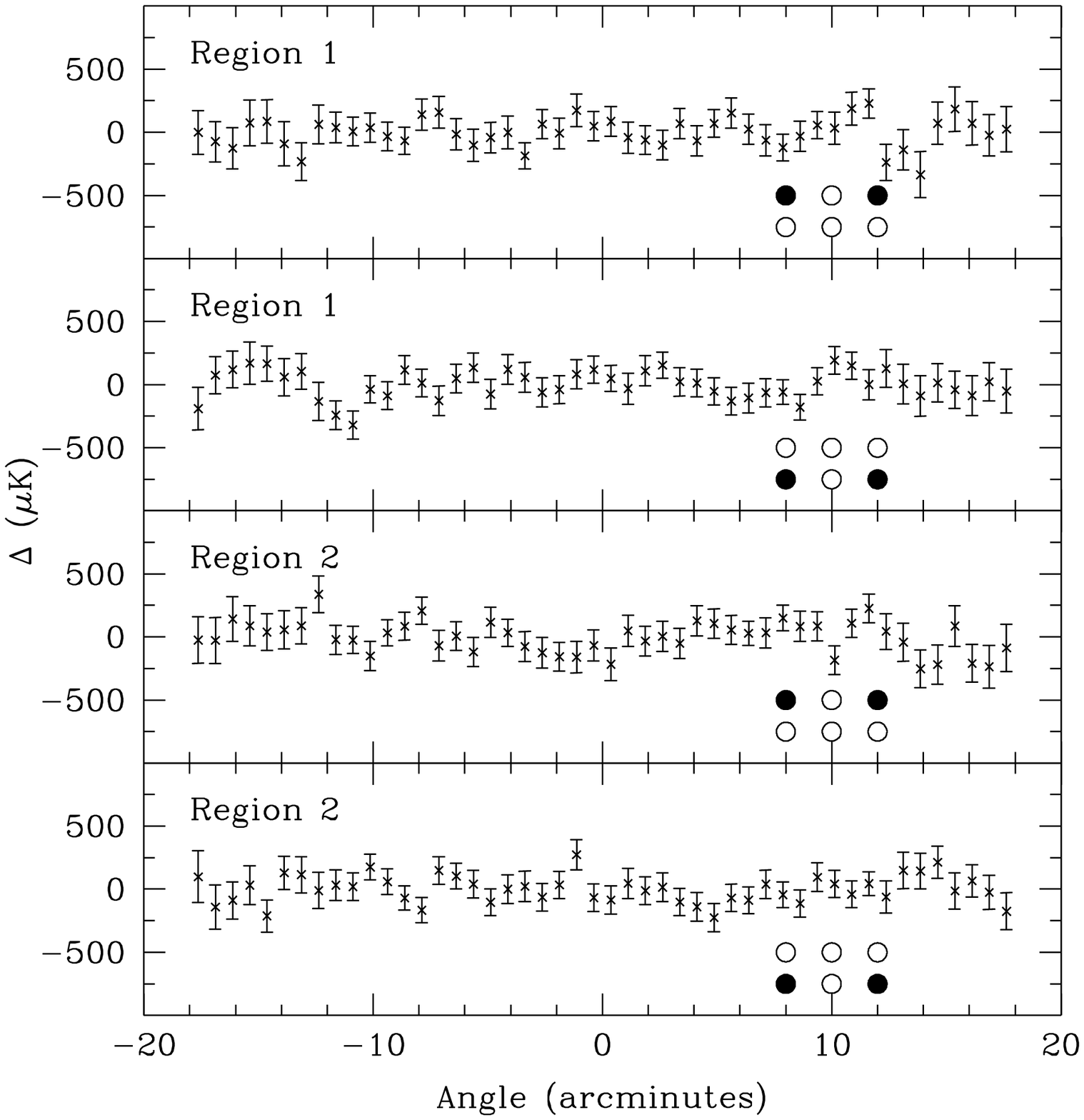}
\end{center}  
\vfill
Figure~\ref{fig:dataplot}

\clearpage
\begin{center}
  \leavevmode
  \epsfxsize=5.5truein
  \epsfbox{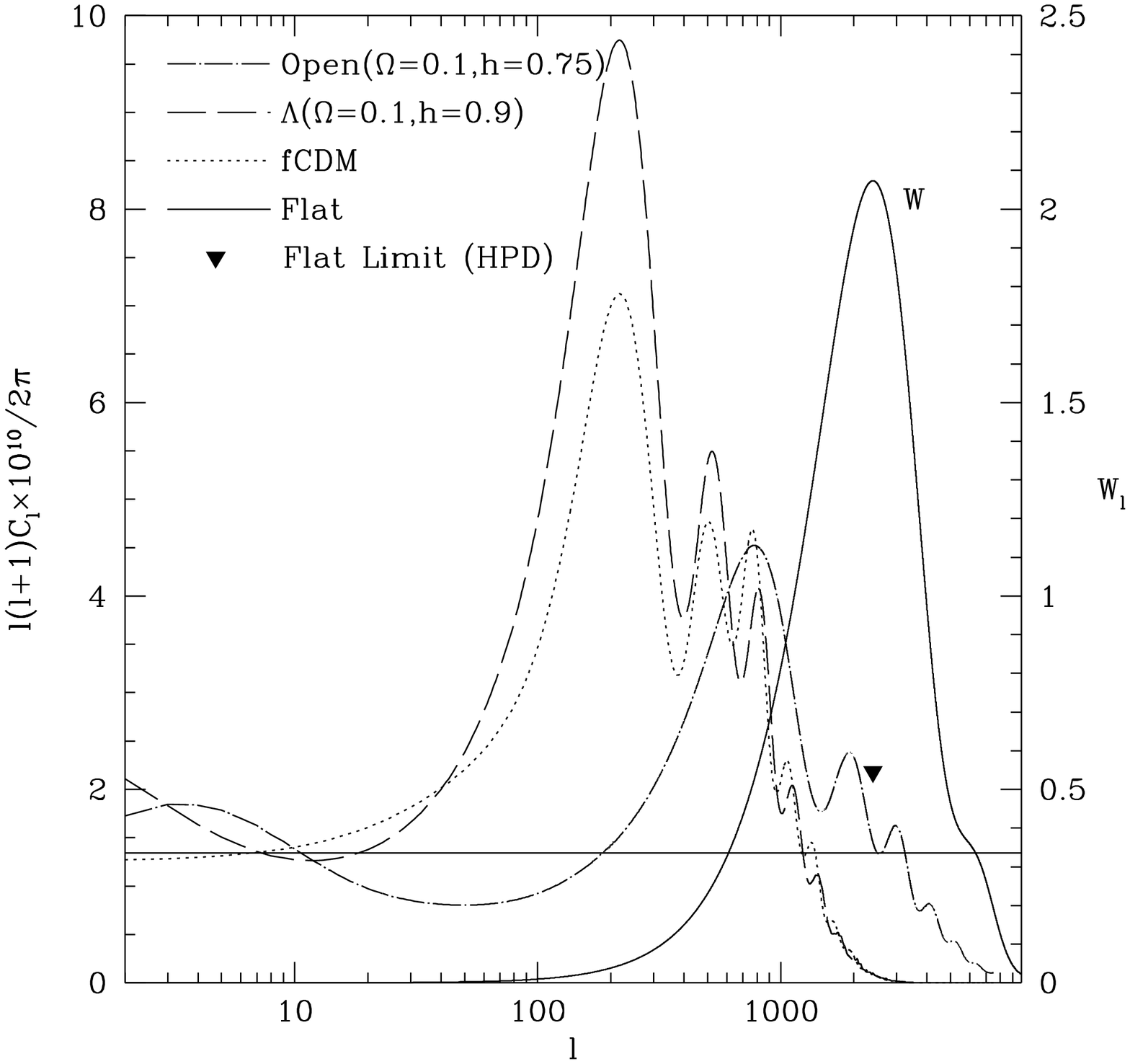}
\end{center}  
\vfill
Figure~\ref{fig:win_mod}

\clearpage
\begin{center}
  \leavevmode
  \epsfxsize=5.5truein
  \epsfbox{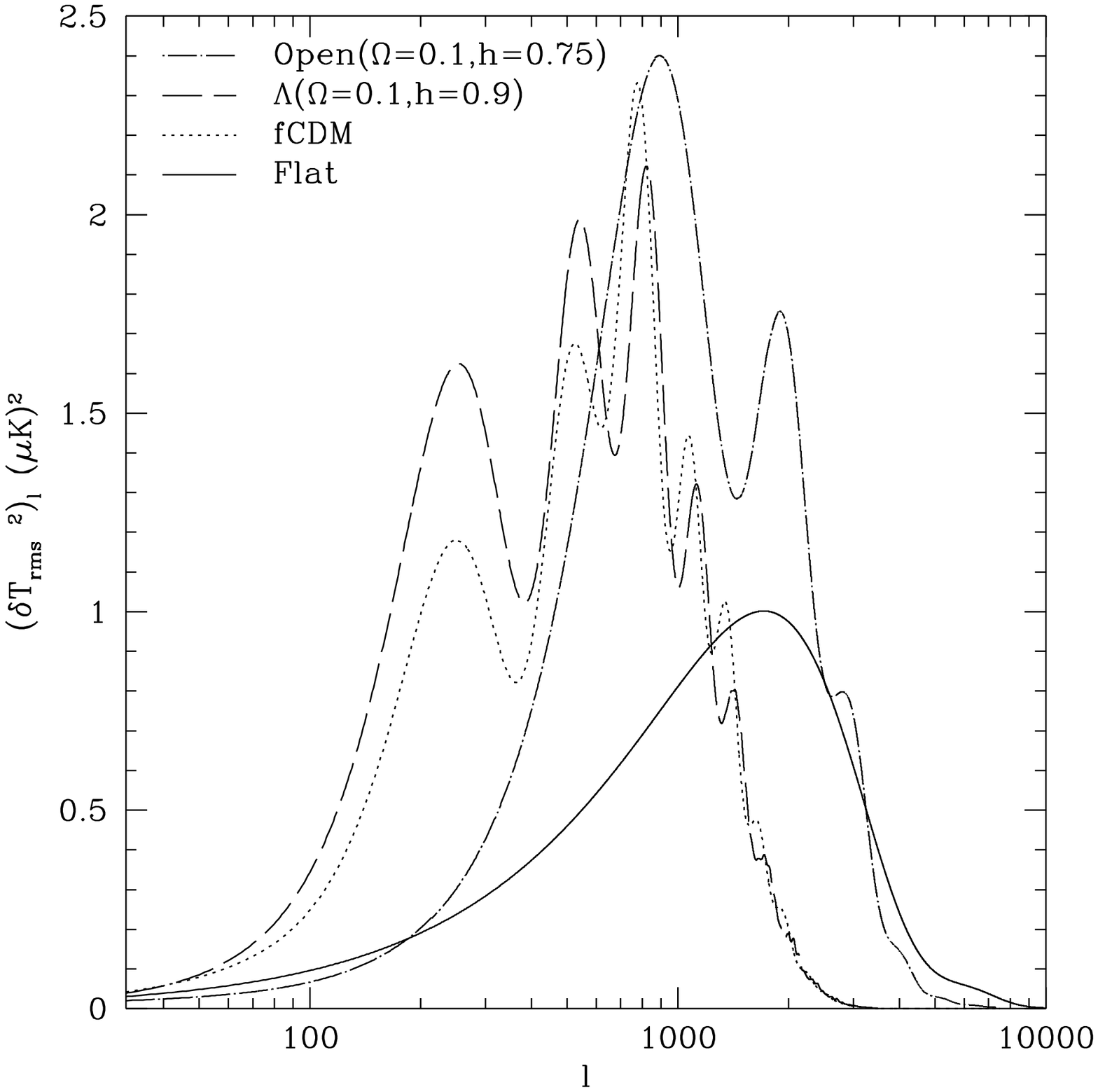}
\end{center}  
\vfill
Figure~\ref{fig:trms}

\clearpage
\begin{center}
  \leavevmode
  \epsfxsize=5.5truein
  \epsfbox{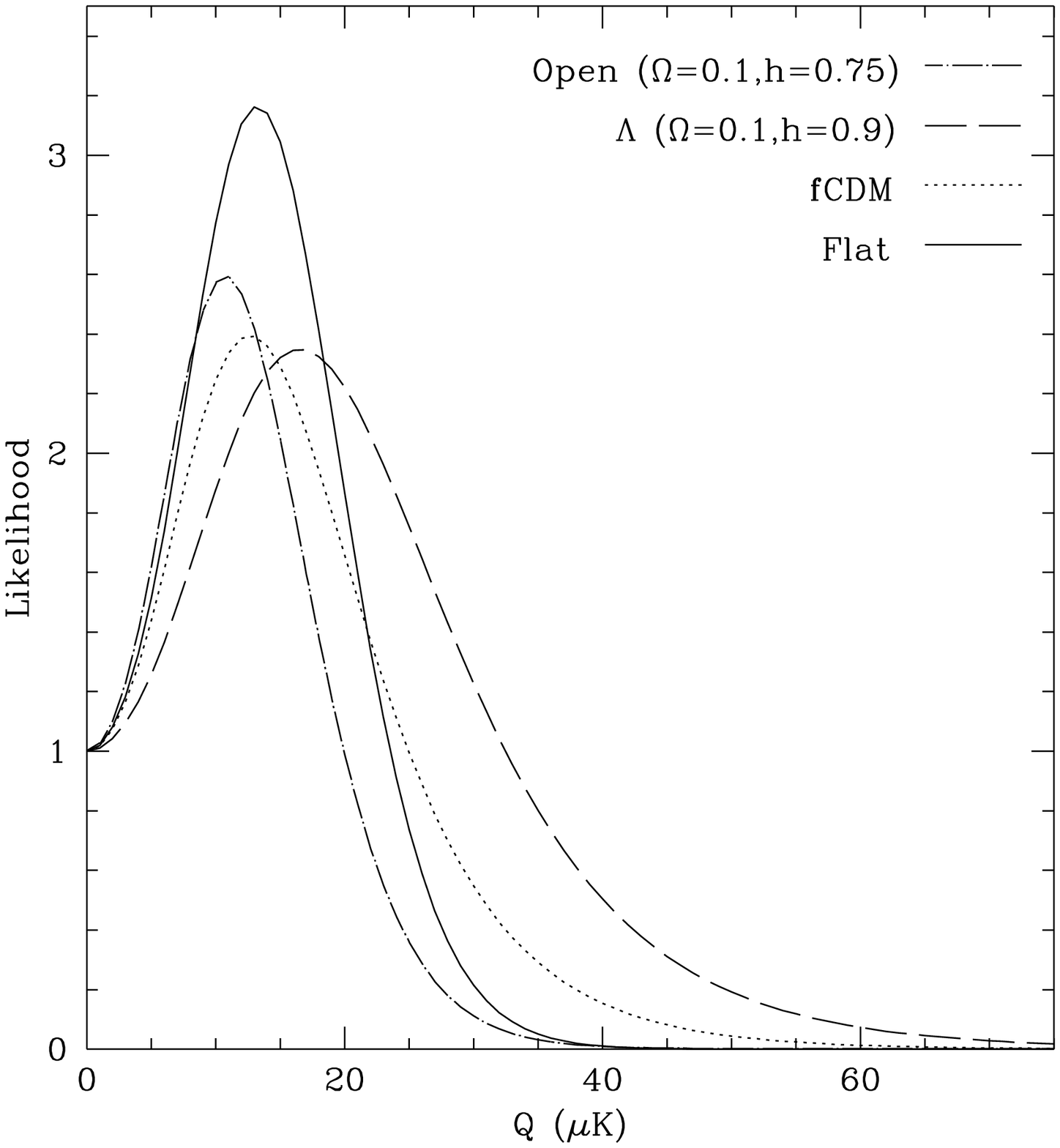}
\end{center}  
\vfill
Figure~\ref{fig:likeplot}

\end{document}

%% file: table1.tex
0.1 & 0.75 & 0.0125 & 37 & 29 & 24 & 42 & 33 & 27 & 40 & 44 \\
0.2 & 0.65 & 0.0175 & 42 & 35 & 28 & 49 & 40 & 31 & 40 & 46 \\
0.2 & 0.70 & 0.0125 & 45 & 38 & 30 & 53 & 44 & 33 & 40 & 46 \\
0.2 & 0.75 & 0.0075 & 50 & 43 & 33 & 58 & 50 & 37 & 40 & 46 \\
0.3 & 0.60 & 0.0175 & 42 & 36 & 27 & 49 & 42 & 31 & 41 & 46 \\
0.3 & 0.65 & 0.0125 & 45 & 40 & 29 & 53 & 46 & 33 & 41 & 46 \\
0.3 & 0.70 & 0.0075 & 51 & 47 & 33 & 60 & 55 & 38 & 41 & 47 \\
0.4 & 0.60 & 0.0175 & 40 & 36 & 26 & 48 & 42 & 30 & 41 & 47 \\
0.4 & 0.65 & 0.0125 & 44 & 41 & 28 & 52 & 48 & 32 & 41 & 47 \\
0.4 & 0.70 & 0.0075 & 50 & 50 & 33 & 60 & 59 & 38 & 42 & 48 \\
0.5 & 0.55 & 0.0175 & 36 & 33 & 23 & 42 & 38 & 26 & 41 & 47 \\
0.5 & 0.60 & 0.0125 & 39 & 38 & 25 & 47 & 44 & 29 & 42 & 48 \\
0.5 & 0.65 & 0.0075 & 46 & 48 & 30 & 56 & 57 & 35 & 43 & 49 \\
\tableline 
1.0 & 0.50 & 0.0125 & 51 & 60 & 35 & 62 & 70 & 41 & 46 & 53 \\
\tableline 
0.1 & 0.90 & 0.0125 & 64 & 72 & 46 & 75 & 82 & 53 & 47 & 55 \\
0.2 & 0.80 & 0.0075 & 66 & 78 & 50 & 77 & 86 & 58 & 50 & 59 \\
0.2 & 0.75 & 0.0125 & 57 & 66 & 40 & 69 & 76 & 46 & 47 & 54 \\
0.2 & 0.70 & 0.0175 & 53 & 57 & 35 & 63 & 67 & 41 & 45 & 52 \\
0.3 & 0.70 & 0.0075 & 61 & 74 & 45 & 73 & 83 & 52 & 50 & 58 \\
0.3 & 0.65 & 0.0125 & 53 & 60 & 36 & 64 & 70 & 41 & 46 & 54 \\
0.3 & 0.60 & 0.0175 & 47 & 51 & 31 & 58 & 60 & 36 & 45 & 52 \\
0.4 & 0.65 & 0.0075 & 59 & 72 & 43 & 71 & 82 & 50 & 50 & 58 \\
0.4 & 0.60 & 0.0125 & 50 & 58 & 34 & 61 & 68 & 39 & 46 & 54 \\
0.4 & 0.55 & 0.0175 & 45 & 48 & 29 & 55 & 57 & 34 & 45 & 52 \\
0.5 & 0.60 & 0.0075 & 51 & 59 & 34 & 62 & 69 & 40 & 46 & 54 \\
\tableline
\multicolumn{3}{c|}{Flat Bandpower} & 36 & 29 & 26 & 41 & 32 & 28 & 40 & 44 \\